# Emergence of Hierarchies in Multi-Agent Self-Organizing Systems Pursuing a Joint Objective

**Gang Chen**

School of Mechanical Engineering

Beijing Institute of Technology, Beijing, P.R. China

No. 5 Zhongguancun South Street, Haidian District, Beijing, China 100081

chengang@bit.edu.cn

**Guoxin Wang**

School of Mechanical Engineering

Beijing Institute of Technology, Beijing, P.R. China

No. 5 Zhongguancun South Street, Haidian District, Beijing, China 100081

wangguoxin@bit.edu.cn

**Anton van Beek**

School of Mechanical and Materials Engineering

University College Dublin, Dublin, Ireland

University College Dublin, Belfield, Dublin 4, Ireland D04 V1W8

anton.vanbeek@ucd.ie

**Zhenjun Ming**[1]

School of Mechanical Engineering

Beijing Institute of Technology, Beijing, P.R. China

No. 5 Zhongguancun South Street, Haidian District, Beijing, China 100081

zhenjun.ming@bit.edu.cn

**Yan Yan**

School of Mechanical Engineering

Beijing Institute of Technology, Beijing, P.R. China

No. 5 Zhongguancun South Street, Haidian District, Beijing, China 100081

yanyan331@bit.edu.cn

---

[1] Corresponding author.

# Emergence of Hierarchies in Multi-Agent Self-Organizing Systems Pursuing a Joint Objective

**Abstract**

In this paper, we address the following question:

*Do hierarchies emerge in teams of agents trained for a joint objective?*

Multi-agent self-organizing systems (MASOS) exhibit key characteristics including scalability, adaptability, flexibility, and robustness, which have contributed to their extensive application across various fields. However, the self-organizing nature of MASOS also introduces elements of unpredictability in their emergent behaviors. This paper focuses on the emergence of dependency hierarchies during task execution, aiming to understand how such hierarchies arise from agents' collective pursuit of the joint objective, how they evolve dynamically, and what factors govern their development. To investigate this phenomenon, multi-agent reinforcement learning (MARL) is employed to train MASOS for a collaborative box-pushing task. By calculating the gradients of each agent's actions in relation to the states of other agents, the inter-agent dependencies are quantified, and the emergence of hierarchies is analyzed through the aggregation of these dependencies. Our results demonstrate that hierarchies emerge dynamically as agents work towards a joint objective, with these hierarchies evolving in response to changing task requirements. Notably, these dependency hierarchies emerge organically in response to the shared objective, rather than being a consequence of pre-configured rules or parameters that can be fine-tuned to achieve specific results. Furthermore, the emergence of hierarchies is influenced by the task environment and network initialization conditions. Additionally, hierarchies in MASOS emerge from the dynamic interplay between agents' "Talent" and "Effort" within the "Environment." "Talent" determines an agent's initial influence on collective decision-making, while continuous "Effort" within the "Environment" enables agents to shift their roles and positions within the system. The insights presented in this paper contribute to a better understanding of self-organizing behaviors and offer guidance for the design and regulation of MASOS.

## Glossary

➢ **Multi-agent self-organizing systems (MASOS)**

MASOS are systems composed of multiple autonomous agents that collaborate in a decentralized manner without relying on a central controller. Each agent operates based on local information and decision-making rules, enabling MASOS to achieve global objectives through distributed collaboration among agents. The inherent self-organizing characteristics of MASOS provide notable advantages, including scalability, adaptability, flexibility, and robustness, thereby facilitating their application across a wide range of domains.

➢ **Emergence**

Emergence refers to the phenomenon in which complex global behaviors arise from the interaction of agents' individual decision-making processes within a system. In the context of MASOS, emergence

> is a fundamental concept, as the global structure or behavior of the system is not explicitly pre-designed but instead arises from the decentralized interactions between individual agents.
>
> ➢ **Hierarchy**
>
> In this paper, hierarchy refers specifically to the dependency hierarchy that emerges among agents during the execution of a joint objective. This hierarchical structure arises from the interdependence between agents, where each agent's actions within the system are functionally dependent on the states of other agents. The dependency hierarchy is quantified by calculating the gradients of agents' actions relative to one another's states. By aggregating these pairwise dependencies, the overall dependency of each agent is determined, offering a metric for evaluating each agent's influence on collective decision-making processes within the system.

# 1 Introduction

Multi-agent self-organizing systems (MASOS) have emerged as a powerful paradigm for tackling complex challenges through the coordinated collaboration of multiple autonomous agents [1, 2]. In various fields, including transportation, logistics, robotics, and manufacturing, MASOS have shown significant advantages in scalability, adaptability, flexibility, and robustness [3-6]. A key characteristic of MASOS is self-organization, where individual agents coordinate their actions and adapt to dynamic environments without any centralized control [7, 8]. Consequently, system-level emergent behaviors and structures arise spontaneously, developing without any hardcoded coordination mechanisms or predetermined organizational schemata.

However, due to the decentralized, autonomous, and partially observable nature of MASOS, the mapping relationship between low-level operational rules and high-level emergent behaviors is inherently nonlinear [9]. This nonlinearity introduces unpredictability in the emergence process [10, 11]. In addition to positive emergent phenomena such as cooperation and optimization, undesirable outcomes (such as collisions, chaos, deadlocks, and failures) may also occur unexpectedly [12, 13]. Therefore, the study of emergence has attracted significant attention from researchers, with extensive efforts aimed at harnessing beneficial emergent behaviors (positive emergence) and mitigating adverse or harmful emergent phenomena (negative emergence) [14]. These efforts are primarily focused on the identification, measurement, classification, control and management of emergent phenomena [15-17].

The emergence of hierarchies, a bottom-up organizational pattern, is widely observed in biological systems such as ant colonies, bee swarms, and wolf packs [18, 19]. These structures are not pre-defined but evolve through spontaneous interactions and self-organization among individual members [20]. Importantly, these hierarchies play a critical role in enhancing a system's performance and capabilities. For example, emergent role specialization within social insects facilitates the development of complex foraging trails and nest construction behaviors [21, 22]. Similarly, human organizations implement hierarchical systems to improve operational efficiency, productivity, and adaptability [23, 24]. In both natural and social systems, hierarchical structures enables collective goal achievement through role differentiation and coordinated actions, thereby substantially enhancing coordination efficiency, decision robustness, and efficient resource allocation [25]. While hierarchical structures have been extensively explored in natural and social systems, whether and how they emerge in MASOS remains unclear. Improving our understanding of this issue is crucial for uncovering the self-organizing behaviors and underlying mechanisms of MASOS.

The objective of this paper is to explore the emergence of hierarchical structures within teams of agents collectively trained for a joint objective. Specifically, we seek to understand the conditions under which hierarchies emerge, how they evolve, and the factors that influence their development. To address this, multi-agent reinforcement learning (MARL) is employed to train MASOS for a collaborative box-pushing task [14]. Three MARL agents are trained to execute a box-pushing task within a two-dimensional, bounded simulation environment that features obstacles and simplified particle dynamics. The interdependence among agents is quantified by calculating the gradients of each agent's actions relative to the states of the other agents. By aggregating these interdependencies, the overall dependency of each agent is derived, providing a metric for its role within the system's hierarchy and facilitating the detection of hierarchical emergence.

The main findings of this paper are summarized as follows:

(1) During the execution of a joint objective task, hierarchical structures emerge within MASOS, and these hierarchies are dynamic, evolving in response to the changing demands of the task. It is noteworthy that these dependency hierarchies emerge spontaneously in response to the joint objective, rather than being artifacts of pre-set rules and parameters that could have been finely tuned to yield specific outcomes. As the task progresses, the hierarchy dynamically adjusts based on the agents' positional advantages and the nature of the task phases, such as linear pushing, objective avoidance, and rotational maneuvers. This adaptability enhances the collective intelligence of the system, as agents take on different roles throughout the task, optimizing their contributions according to real-time task requirements.

(2) The emergence of the hierarchy is influenced by both the task environment and network initialization conditions. As task settings change (e.g., target position or obstacle configurations), the dependency patterns evolve accordingly. This reorganization prioritizes agents with positional advantages in specific task phases, with those agents playing a central role in the respective phases. Additionally, varying network initialization conditions lead to different hierarchical structures: MASOS may exhibit persistent dominance, where a single agent consistently leads, or alternating dominance, where leadership shifts depending on task phases.

(3) The hierarchy emerges from the dynamic interplay between the agents' "Talent" and "Effort" within the "Environment" during task execution. An agent' "Talent," such as its positional advantage or favorable network initialization conditions, combined with its "Effort," represented by network updates achieved through learning, interact to determine the agent' influence on a team's collective performance. While "Talent" sets the starting point, continuous "Effort" within the "Environment" allows agents to shift their roles and positions within the system. This interplay enables MASOS to develop a dynamically evolving hierarchy during task execution.

The remainder of this paper is structured as follows. A review of the relevant literature is provided in Section 2. The research method is presented in Section 3. A detailed case study of a box-pushing problem is introduced in Section 4. The results and discussion are detailed in Section 5. Finally, the conclusion and potential directions for future work are outlined in Section 6.

## 2 Literature Review

In this section, we outline the fundamental concepts and characteristics of MASOS. Subsequently, we review key studies on emergence, with emphasis on the emergence of hierarchical structures. Finally, we will introduce the MARL method, highlighting its application in studying MASOS. The key characteristics of the selected papers on MASOS and emergence are summarized in Table 1. Specifically, this table provides a comparative overview of various studies, highlighting the emergence dynamics investigated, the research methods employed, and the case studies addressed. The emergence dynamics in these studies primarily explore various aspects of emergent behaviors, including their identification, classification, measurement, and management. Notably, some studies [20, 25] examine the emergence of hierarchy, investigating how hierarchical structures form and evolve through agent interactions during task execution. The research methods employed across these studies are diverse, including rule-based method (RBM), agent-based modeling (ABM), reinforcement learning (RL), and MARL. The case studies presented in these works cover a broad range of applications, including distributed task allocation, collaborative scheduling, generative design, complex assembly tasks, and a box-pushing problem. A systematic review of the relevant literature is presented in Sections 2.1 to 2.3.

Table 1. Summary and comparison of related works

| Literature | Emergence dynamics | | | | | Research method | Case study |
|---|---|---|---|---|---|---|---|
| | Identification | Classification | Measurement | Management | Hierarchy | | |
| Ji et al. [26-28] | ✓ | | ✓ | | | MARL | Box-pushing problem |
| Huang et al. [7, 8] | / | / | / | / | / | RL | Complex assembly tasks (box-pushing problem) |
| Su et al. [29] | / | / | / | / | / | RBM | Generative design |
| Ming et al. [9] | ✓ | | ✓ | ✓ | | RBM | Box-pushing problem |
| Jiang et al. [14] | ✓ | | ✓ | ✓ | | MARL | Box-pushing problem |
| Hejazi et al. [30] | | | | | ✓ | MARL | Distributed task allocation |
| Han et al. [31] | ✓ | | | | | ABM | Multi-agent game model |
| Li et al.[32] | ✓ | ✓ | ✓ | | | ABM | Collaborative scheduling |
| Singh et al. [15] | ✓ | ✓ | | | | ABM | Swarms of unmanned aerial vehicles |
| Sharma et al. [16] | ✓ | | | | | RL | Target acquisition tasks |
| Grupen et al. [17] | ✓ | | ✓ | | | MARL | Collaborative cooking |
| Kalantari et al. [10, 12, 13] | ✓ | ✓ | ✓ | ✓ | ✓ | RBM | NASA Autonomous Nano Technology Swarm mission |
| Chen et al. [33] | ✓ | ✓ | | | ✓ | ABM | Software development, consulting, and Minecraft game |
| Ohnishi et al. [20] | ✓ | | ✓ | | ✓ | ABM | Fish shoal and flying bird flock |
| Deffuant et al. [25] | ✓ | | ✓ | | ✓ | ABM | Simulations of opinion evolution among groups |
| Hahn et al. [34] | ✓ | | ✓ | | | MARL | Predator-prey pursuit game |
| Gui et al. [35] | / | / | / | / | / | MARL | Collaborative scheduling |
| Li et al. [5, 6] | / | / | / | / | / | MARL | Collaborative scheduling |
| Martinez-Gil et al. [36] | ✓ | | ✓ | | | MARL | Pedestrian systems |
| Our work | ✓ | | ✓ | | ✓ | MARL | Box-pushing problem |

## 2.1 MASOS

MASOS are systems composed of multiple autonomous agents collaborating in a decentralized manner without relying on a single central controller [26-28]. Each agent operates based on local information and decision rules, enabling adaptive responses to changes in the external environment. Through iterative feedback among these numerous local interactions, MASOS collectively generate complex global behaviors that cannot be directly inferred through simple aggregation of individual agents' decision-making processes [37, 38]. Unlike traditional centralized systems, where a central controller directs the behavior of all agents, MASOS can accomplish overarching tasks or achieve global objectives through distributed collaboration in complex scenarios.

The self-organizing characteristics of MASOS confer distinctive advantages in terms of scalability, adaptability, flexibility, and robustness, which have facilitated their widespread application across diverse domains [3, 4]. For example, Huang et al. [7, 8] applied MASOS to complex assembly tasks using an "L-shape" assembly task as a testbed. They investigated the impact of reward shaping on both the learning process and overall task performance and further examined the role of social learning within these systems. Su et al. [29] employed a Monte Carlo tree search-based MASOS to address generative design challenges in complex floorplans for high-rise residential buildings, leveraging this hybrid approach to efficiently explore multi-objective layout solutions, thereby enhancing overall design flexibility and quality. Ming et al. [9] and Jiang et al. [14] applied MASOS to a box-pushing problem, optimizing time efficiency, energy efficiency, and system reliability. They utilized surrogate models and MARL, presenting innovative approaches to the design of self-organizing systems. Hejazi et al. [30] employed MASOS to tackle the distributed task allocation problem, focusing on how to identify both the optimal communication structure and the optimal task strategy within these systems. Moreover, MASOS have also been explored in collaborative scheduling [5], collaborative navigation [39], and disaster relief [40].

Despite the promise of MASOS, several core challenges must be addressed to enable widescale deployment. Because of the decentralized and autonomous nature of MASOS, the relationship between low-level rules and high-level emergent performance is highly non-linear [41]. This non-linearity often leads to unpredictable system behaviors, including negative emergent phenomena such as collisions, chaos, deadlocks, and failures, which are difficult to be anticipated by the designers [9]. Consequently, improving our understanding of emergent behaviors in MASOS is critical not only for directing systems toward optimal performance but also for preventing adverse effects.

## 2.2 Emergence

Emergence refers to the phenomenon in which complex global behaviors arise from the interaction of agents' individual decision-making processes within a system [10]. Emergence manifests in different forms (positive/negative) and shapes (types) across various systems [41]. Positive emergence can be harnessed to achieve efficient task allocation, communication, and decision-making in highly distributed and uncertain environments, whereas negative emergence may result in adverse outcomes such as system instability, chaotic dynamics, and operational failures [10]. In MASOS, emergence is a key concept, as the global structure or behavior of the system is not explicitly pre-designed but instead emerges from the decentralized interactions among individual agents [13]. The investigation of emergent behaviors in MASOS is essential for understanding how individual agents, without the presence of a central controller, can coordinate effectively to achieve complex, collective tasks.

Emergent behaviors in MASOS cannot be ascribed to any individual agent but instead result from the coordination and interactions among agents, manifesting as a collective effort [31]. These behaviors typically emerge from local interactions governed by agent-specific rules. Such interactions facilitate the emergence of global structures, task allocation, and collective decision-making processes [32]. In contrast to traditional systems, where outcomes are predetermined by a central authority, the behaviors in MASOS emerge spontaneously and are often characterized by their unpredictability [12]. This characteristic of emergent behaviors is a centralizing aspect of MASOS research, as it challenges conventional notions of system behavior by demonstrating that "the whole is greater than the sum of its parts." The unpredictability of emergent behaviors presents significant challenges in controlling and optimizing MASOS. The complexity of emergent behaviors often leads to system unpredictability posing safety risks or leading to system failures [12].

In addition to clarifying the definitions and characteristics of emergence, substantial research efforts have focused on the identification, classification, measurement, and management of emergent behaviors, attracting significant attention from researchers [10]. For instance, Singh et al. [15] proposed a multi-agent simulation framework to identify and classify emergent behaviors. Their approach used agent-based modeling to identify how local interactions among agents led to the emergence of global behaviors and to classify those behaviors according to Fromm's taxonomy [42]. Sharma et al. [16] focused on the identification of emergent behaviors among autonomous agents in target acquisition tasks. They constructed spatio-temporal heatmaps of the agents' positional trajectories, extracted key feature sets that capture underlying behavioral regularities, and employed Principal Component Analysis and clustering to distinguish emergent behavior patterns. Grupen et al. [17] addressed the classification and measurement of emergent behaviors in multi-agent systems using a concept-based approach. By conditioning each agent's action on human-understandable concepts, their approach enables post-hoc behavioral analysis through concept intervention, revealing the mechanisms underlying agent collaboration and identifying lazy agents (i.e., those that fail to contribute to team reward through their individual actions). Kalantari et al. [13] proposed an entropy-based, goal-oriented approach for the management of emergent behaviors in self-organizing systems. They leveraged a feedback control loop to dynamically adjust system parameters based on real-time entropy measures, thereby enhancing the coordination and efficiency of emergent behaviors as demonstrated in the NASA Autonomous Nano Technology Swarm mission [43].

In MASOS, social behaviors analogous to those observed in human societies can spontaneously arise during collaboration, ranging from beneficial phenomena such as volunteer behavior and conformity behavior to potentially harmful destructive behavior [33]. Given the shared characteristics of group collaboration in human teams, animal groups, and agent teams, hierarchical structures observed in human social activities and animal groups may also emerge in MASOS [20]. Importantly, these hierarchical structures are not externally imposed but emerge spontaneously through interactions among individuals and collective alignment of opinions. Minor initial differences are amplified over repeated interactions, leading to stable hierarchical rankings [25].

The focus of this paper is on the emergence of hierarchy in MASOS. Specifically, we aim to explore the following research question: Do teams of agents trained for a joint objective naturally develop hierarchical structures? Investigating this question will shed light on the self-organizing mechanisms and evolutionary dynamics of multi-agent systems.

## 2.3 MARL

MARL refers to the extension of RL to environments involving multiple agents that interact with each other and the environment to achieve their individual or joint objectives [44]. Unlike traditional RL, where an agent learns in isolation, MARL involves learning strategies in environments where agents' actions influence one another, leading to complex interdependencies and interactions [45, 46]. In MARL, each agent seeks to optimize its own policy based on the rewards it receives, which are typically dependent on the actions taken by other agents in the environment [47]. Agents must learn to balance exploration (trying new actions to gain knowledge) and exploitation (leveraging known actions that yield high rewards) [6].

Owing to its decentralized decision-making, distributed coordination, and adaptive learning capabilities, MARL has emerged as a key approach for investigating MASOS [48, 49]. In MASOS, agents must collaborate or compete to achieve global objectives without relying on a centralized controller, which aligns closely with the core principles of MARL [5]. Hahn et al. [34] explored the emergence of flocking behavior in a scenario where multiple autonomous agents (prey) were trained using MARL to evade predator capture. Their study revealed that interactions among self-interested agents can spontaneously generate collective behaviors. This demonstrates the potential for MARL to simulate emergent, adaptive behaviors without explicit programming.

In the literature, MARL is extensively applied to model and optimize collaborative tasks within MASOS. For example, Gui et al. [35] investigated a self-organizing manufacturing system employing MARL to facilitate collaborative dynamic scheduling, thereby enhancing coordination and operational efficiency in highly dynamic manufacturing environments. Li et al. [5] proposed an innovative scheduling approach that integrates multi-agent systems with MARL. In this approach, manufacturing resources are modeled as autonomous agents with self-organizing capabilities, and these agents utilize MARL algorithms to learn optimal scheduling strategies through interactions and experiences within the manufacturing environment. Martinez-Gil et al. [36] explored the efficacy of MARL in capturing emergent behaviors in pedestrian systems. Their study specifically examined how local interactions among individual agents give rise to collective phenomena, such as lane formation, crowd segmentation, and effective collision avoidance, thereby advancing the understanding of complex, self-organizing behaviors in multi-agent environments.

In this paper, we leverage MARL to establish MASOS, thereby enabling a team of agents to execute a box-pushing task. The focus is on investigating whether hierarchical structures naturally emerge in MASOS and to study the implications of those structures when the agents collectively pursue a joint objective.

## 3   Research Method

In this section, we address the following research question using the overall research framework depicted in Fig. 1: *Do hierarchies emerge in teams of agents trained for a joint objective?* First, we present the MARL algorithm employed in this paper (Section 3.1), with a focus on the framework of centralized training with decentralized execution (CTDE) for training a team of agents. Subsequently, we analyze how the system structure critically determines system performance (Section 3.2). Finally, we present the method utilized for identifying the emergence of dependency hierarchies within the system (Section 3.3).

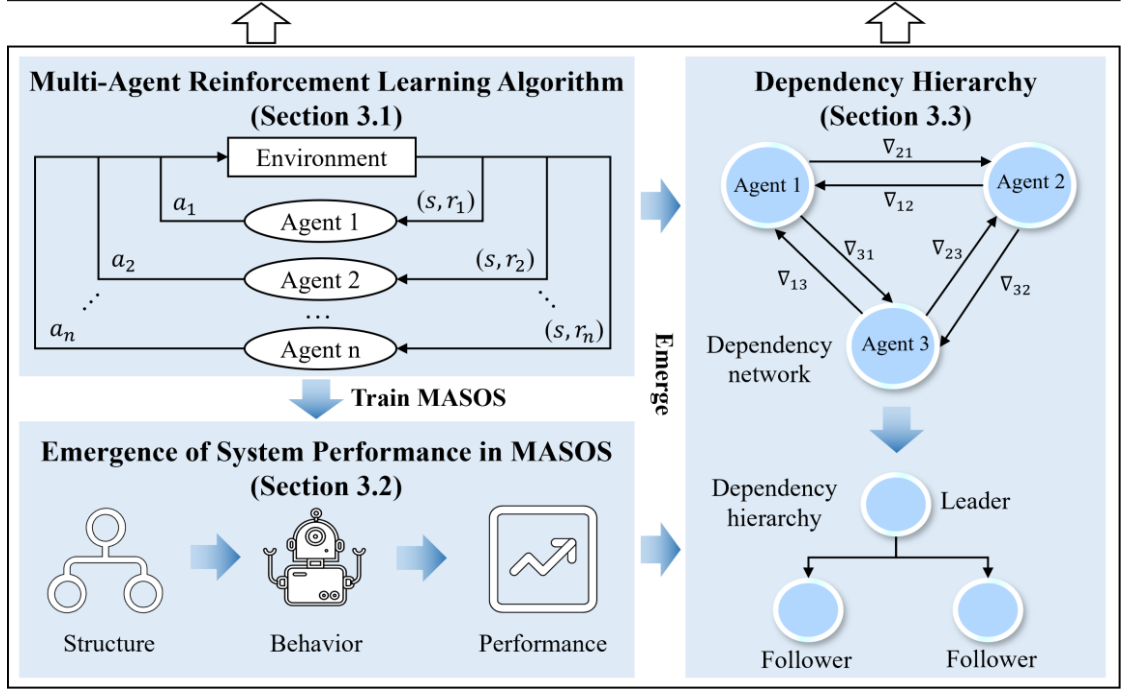

Fig. 1 Overall research framework: hierarchy emergence in teams of agents trained for a joint objective

## 3.1 MARL Algorithm

MARL enables agents to iteratively refine their decisions through trial-and-error interactions, thereby replicating the dynamic emergence processes in MASOS while simultaneously identifying critical factors that influence system behavior and enhance overall performance. Therefore, in this paper, MARL is employed to train agents within the system, to effectively investigate the mechanisms underlying emergent behaviors and improving the operational efficiency and adaptability of MASOS.

To implement the training of agents, we adopt CTDE [50, 51], the typical MARL framework illustrated in Fig. 2. In this framework, each agent is equipped with an individual actor-critic structure. The actor-network of each agent receives its local observation as input and outputs a corresponding action to interact with the environment. Simultaneously, the critic-network evaluates the value of actions using global state information to calculate the action-value function [52]. During training (green region in Fig. 2), agents share information through centralized learning, allowing the critic-networks to incorporate global information for more accurate evaluation and updating of the actor-networks. However, during execution (orange region in Fig. 2), each agent relies solely on its local observation and actor-network to make decisions, ensuring decentralized execution while maintaining coordination within the system [53]. This design effectively leverages the advantages of both centralized training and decentralized decision-making, enabling efficient and adaptive control in MASOS.

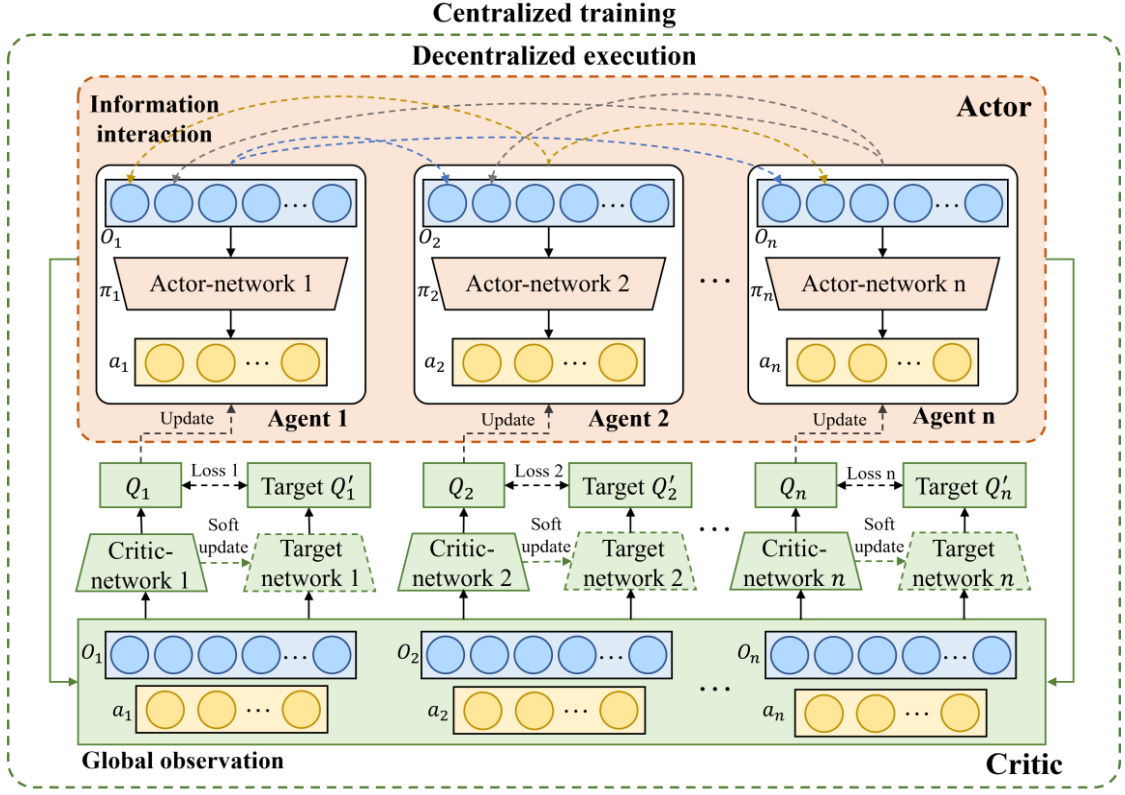

Fig. 2 Framework of centralized training with decentralized execution

Building upon the CTDE framework, we employed the Multi-Agent Deep Deterministic Policy Gradient (MADDPG) algorithm [50]. MADDPG is an extension of the Deep Deterministic Policy Gradient (DDPG) algorithm [54], specifically designed for multi-agent environments. Its deterministic policy output and ability to handle continuous action spaces make it particularly advantageous for stable and efficient learning in multi-agent settings [35]. For each agent $i$ in a system with $n$ agents, the current parameters of the actor-networks and critic-networks are denoted as $\theta^{a_i}$ and $\theta^{c_i}$, respectively, where $i = 1, \dots, n$. In addition, the corresponding target network parameters are denoted as $\theta^{a_i'}$ and $\theta^{c_i'}$. At time step $k$ of task execution (i.e., a discrete decision point), each agent $i$ interacts with the environment through its actor network, using its local observation $O_{i,k}$, and selects an action $a_{i,k}$ according to its policy $\pi_i$:

$$a_{i,k} = \pi_i(O_{i,k}|\theta^{a_i}), \quad i = 1, \dots, n, \quad (1)$$

where $O_{i,k}$ denotes the observation collected by agent $i$ at time step $k$, and $\pi_i$ is the policy network, parameterized by $\theta^{a_i}$, that maps observations to actions.

The environment is represented by a global state variable $s_k$, and the joint action taken by all agents is given as $a_k = (a_{1,k}, \dots, a_{n,k})$. Executing $a_k$ in state $s_k$ transitions the environment to a new state $s_{k+1}$, and each agent receives a corresponding reward $r_{i,k}$. The transition tuple $(s_k, a_k, r_{i,k}, s_{k+1})$ is stored in the shared experience replay buffer.

The critic-network for agent $i$ is trained to approximate the action-value function $Q_i(s, a)$. Given a mini-batch of $N$ samples indexed by $j \in \{1, \dots, N\}$, the target value for each sample $j$ is computed using the target networks:

$$y_j = r_{i,j} + \gamma Q_i^{target}(s_j', a_1', \dots, a_n'), \quad i = 1, \dots, n, \ j = 1, \dots, N, \quad (2)$$

where $a_j' = \pi_i^{target}(O_j')$ denotes the action generated by the target actor-networks and $\gamma$ is the discount factor.

The critic loss is defined as the mean squared error between the predicted and target Q-values:

$$L(\theta^{c_i}) = \frac{1}{N}\sum_{j=1}^{N}\left(Q_i(s_j, a_j) - y_j\right)^2, \quad i = 1, \dots, n, \tag{3}$$

where $a_j = (a_{1,j}, \dots, a_{n,j})$ is the joint action in sample $j$. This loss function ensures the critic-network learns to approximate the expected Q-values based on the reward and future Q-value estimates.

The actor-network is updated by maximizing the expected Q-value, while the actions of all other agents are held fixed:

$$J(\theta^{a_i}) = \frac{1}{N}\sum_{j=1}^{N} Q_i\left(s_j, a_1^j, \dots, \pi_i(o_i^j), \dots, a_n^j\right), \quad i = 1, \dots, n. \tag{4}$$

To ensure training stability, the parameters of the target networks (both actor and critic) are updated using a soft update mechanism:

$$\begin{cases} \theta^{a_i'} \leftarrow \tau\theta^{a_i} + (1-\tau)\theta^{a_i'} \\ \theta^{c_i'} \leftarrow \tau\theta^{c_i} + (1-\tau)\theta^{c_i'} \end{cases}, \quad i = 1, \dots, n, \tag{5}$$

where $\tau \in (0,1)$ is the soft update coefficient.

To implement the policy described above, each agent selects its actions based on a comprehensive set of observation variables that extend beyond its own state to include information about the environment, the task objectives, and the status of other team members. We categorize these observations into four types:

1) Agent-related variables: Capturing the agent's own dynamic state—such as position and velocity—which serve as primary inputs for its decision-making.

2) Environment-related variables: Describing external features—such as the location, shape, and size of obstacles—to support collision avoidance and feasible path planning.

3) Task-related variables: Specifying task objectives—such as target point locations—so that agents can adapt their behavior to meet current goals.

4) Team-related variables: Reflecting the states and actions of other agents—such as their positions and velocities—to enable coordination, task allocation, and conflict avoidance within the team.

## 3.2 Emergence of System Performance in MASOS

The emergence of system performance in MASOS is driven by its underlying implicit structure. In MASOS, individual agents typically interact based on localized rules. Although each agent's behavior is local, specific structural arrangements such as hierarchical organization enable local interactions to collectively influence global performance. For instance, inter-agent dependencies and collaborative patterns among agents lead to emergent collective behaviors, ultimately manifesting as system-level optimization or adaptive capabilities. Furthermore, structures of MASOSs often exhibit high adaptability and evolutionary capacity, allowing them to dynamically adjust in response to environmental changes and agent interactions. This structural adaptability serves as the foundation for continuous system optimization and performance enhancement.

Inspired by the widely used function-behavior-structure process in product design [55, 56], and recognizing its relevance to the design of MASOS [57], we apply this framework to MASOS. In MASOS, the system structure governs the interactions between agents, the agents' behaviors emerge based on these interactions, and the function is measured by the system's performance in achieving its goals. The "structure-behavior-performance" mechanism within MASOS is depicted in Fig. 3, illustrating how the system structure shapes agent behaviors, ultimately facilitating the emergence of performance. Here, structure refers to the organizational framework governing agent interactions, including connectivity patterns, task allocation, and resource-sharing mechanisms. Structures may either be static or dynamically adjustable, determining how agents interact, how information flows, and how tasks and resources are distributed. Behavior represents the actions taken by agents based on their perception of the environment and their goals within the given structure. The structure defines the interaction protocols and decision-making processes between agents, while behavior reflects the practical execution of these protocols. Performance, as the ultimate system output, is typically assessed by whether the system achieves its predefined goals, such as task completion, operational efficiency, and optimal resource utilization.

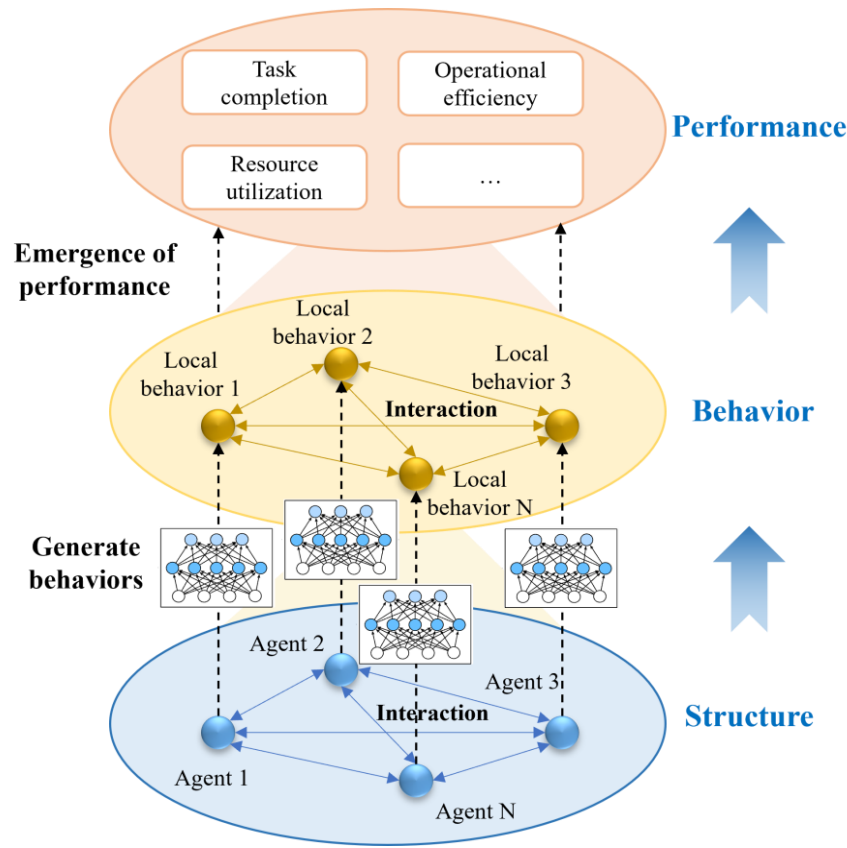

Fig. 3 The "structure-behavior-performance" mechanism within MASOS

The structure governs the interactions, collaborations, and task distributions among agents; behavior reflects the agents' responses and interactions shaped by this structure; and performance emerges as the aggregate outcome of agent behaviors, facilitated by global cooperation to achieve system objectives. Therefore, understanding the emergence of structure in MASOS and its underlying mechanisms is essential for enhancing overall system performance.

### 3.3 Dependency Hierarchy

In MASOS, the dependency hierarchy plays a critical role in determining the system's performance. The collaboration and information flow among agents are inherently dependent on this hierarchical structure. For example, task allocation, resource sharing, and information transmission are all influenced by these inter-agent dependencies. Formally, this can be represented as a directed graph, where nodes correspond to individual agents, and directed edges (e.g., A→B) signify that agent B's actions are functionally dependent on agent A's state.

In the MARL algorithm, each agent generates its actions based on observation variables that include not only its own state information but also the states of other agents. Therefore, each agent's actions depend on the states of other agents in the system [58]. This dependency can be quantified by computing the gradient of an agent's actions with respect to other agents' states, where a larger gradient magnitude indicates greater sensitivity and consequently stronger dependency between agents. In MASOS, these dependency relationships are typically mutual and bidirectional. The complete set of inter-agent dependencies forms a dependency network, as illustrated on the left of Fig. 4. The aggregation of these pairwise dependencies yield a net dependency value for each agent, representing its overall influence within the system. When an agent's net dependency value is the highest among all agents, it fundamentally demonstrates that the agent exerts the most critical influence in collective decision-making processes and may even assume something akin to leadership responsibilities. In MASOS working towards a joint objective, agents are then ranked by their net dependency values and grouped into different roles (e.g., leaders and followers), thereby giving rise to distinct dependency hierarchies, as illustrated on the right of Fig. 4. These dependency hierarchies evolve dynamically in response to shifting inter-agent dependencies throughout task execution.

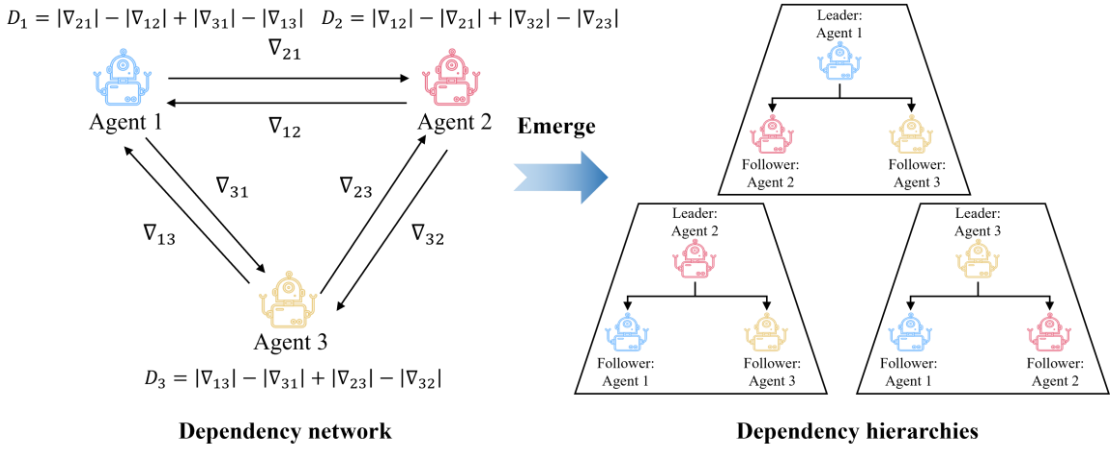

Fig. 4 Emergence of dependency hierarchy

The dependency between agents is quantified by computing the gradient of an agent's actions with respect to the states of other agents. Specifically, a larger gradient of agent $i$'s action relative to agent $j$'s state indicates higher sensitivity. This demonstrates that the behavior of agent $i$ is more dependent on the information from agent $j$, reflecting a more significant influence of agent $j$ on agent $i$'s decision-making process. Each agent generates its action through its policy network $\pi_i$ based on the joint observation space. Formally, the action of agent $i$ is determined by

$$a_i = \pi_i(O_1, O_2, \ldots, O_n, O_{other}|\theta_i), \quad i = 1, \ldots, n, \tag{6}$$

where $O_i, i = 1, \ldots, n$ represents the observation variables related to the state of agent $i$, and $O_{other}$ includes other observation information related to the environment and the task.

To quantify directional dependencies, we compute the sensitivity of agent $i$'s action to agent $j$'s state through the gradient operator. Formally, the gradient of agent $i$'s action $a_i$ with respect to agent $j$'s state observation $O_j$ is defined as

$$\nabla_{ij} = \nabla_{O_j} a_i \triangleq \frac{\partial \pi_i(O_1, \ldots, O_n, O_{other}|\theta_i)}{\partial O_j} \quad i, j = 1, \ldots, n, \tag{7}$$

where $\nabla_{O_j}$ denotes the gradient operator applied to $O_j$, $a_i = \pi_i(\cdot)$ represents agent $i$'s action as defined in Eq. (6), and the partial derivative explicitly shows the functional dependence of $\pi_i$ on $O_j$.

The gradient $\nabla_{ij}$ measures the sensitivity of agent $i$'s action to the state information from agent $j$, highlighting the conditional dependency between agents. The magnitude of $\nabla_{ij}$ directly indicates the strength of this directional dependence, with larger values signifying stronger behavioral reliance of agent $i$ on agent $j$'s state. By aggregating bidirectional dependencies between agents, we obtain each agent's net dependency measure, which quantifies its system-level influence during cooperative task execution. The dependency $D_i$ of agent $i$ is calculated as

$$D_i = \sum_{j \neq i} (|\nabla_{ji}| - |\nabla_{ij}|) \quad i,j = 1, \dots, n, \tag{8}$$

where $|\nabla_{ji}|$ represents the dependence of agent $j$ on agent $i$, while subtracting $|\nabla_{ij}|$ accounts for the reverse influence from agent $j$ to agent $i$. This calculation makes sure $D_i$ accurately reflects the net dependency of agent $i$ over other agents in the system.

For instance, in a team of three agents (as illustrated in Fig. 4), the formula for calculating the dependency value of each agent is given as

$$\begin{cases} D_1 = |\nabla_{21}| - |\nabla_{12}| + |\nabla_{31}| - |\nabla_{13}|, \\ D_2 = |\nabla_{12}| - |\nabla_{21}| + |\nabla_{32}| - |\nabla_{23}|, \\ D_3 = |\nabla_{13}| - |\nabla_{31}| + |\nabla_{23}| - |\nabla_{32}|. \end{cases} \tag{9}$$

The emergence of hierarchical structures in MASOS during task execution can be quantitatively assessed through the analysis of these dependency values. Specifically, agents with varying dependency values evolve into distinct roles within the team. Agents with higher dependency values exert greater influence on the system, as they are more prominently considered by their peers during collaborative tasks, thereby significantly impacting team decision-making processes. Typically, agents with higher dependency values assume leadership roles (leaders), while those with lower dependency values take on subordinate roles (followers). The pseudocode for identifying the emergence of hierarchical structures in MASOS is presented in Table 2.

Table 2. Pseudocode for identifying the emergence of hierarchies in MASOS

| **Identify the emergence of hierarchies in MASOS** |
|---|
| 1:  // Agents output actions |
| 2:  **for** $i = 1$ **to** $n$ **do** |
| 3:      $a_i \leftarrow \pi_i(O_1, O_2, \dots, O_n, O_{other}|\theta_i)$   // Eq. (6) |
| 4:  **end for** |
| 5:  // Compute gradients of agents' actions with respect to other agents' states |
| 6:  **for** $i = 1$ **to** $n$ **do** |
| 7:      **for** $j = 1$ **to** $n$ and $j \neq i$ **do** |
| 8:          $\nabla_{ij} \leftarrow \nabla_{O_j} a_i \triangleq \frac{\partial \pi_i(O_1, \dots, O_n, O_{other}|\theta_i)}{\partial O_j}$   // Eq. (7) |
| 9:      **end for** |
| 10: **end for** |
| 11: // Compute dependency value for each agent |
| 12: **for** $i = 1$ **to** $n$ **do** |
| 13:     $D_i \leftarrow \sum_{j \neq i}(|\nabla_{ji}| - |\nabla_{ij}|)$   // Eq. (8) |
| 14: **end for** |
| 15: // Identify the emergence of hierarchies |
| 16: **for** $i = 1$ **to** $n$ **do** |
| 17:     **for** $j = 1$ **to** $n$ **do** |
| 18:         **if** $D_i > D_j$ and $j \neq i$ **then** |
| 19:             There is an emergence of dependency hierarchies. |
| 20:             agent $i$ is a leadership role (leader) and agent $j$ is a subordinate role (follower). |
| 21:         **else** |
| 22:             There is not emergence of hierarchies. |
| 23:         **end if** |
| 24:     **end for** |
| 25: **end for** |

# 4 Illustrate Example: A Box-Pushing Problem

In this section, we explore the emergence of hierarchies in MASOS in a box-pushing exercise [59]. The box-pushing task is a focal problem in multi-agent systems, where multiple agents collaborate through coordinated actions to push a box and thereby execute a series of movements [60-62]. The simulation environment is built upon the OpenAI Gym Multi-agent Particle Environment (MPE) repository and developed for use with MASOS [50, 52].

## 4.1 Task Description

The box-pushing task involves multiple agents working collaboratively to push a box toward a target position while avoiding obstacles. The target position varies in each scenario, being located at the top-left, top-right, or directly above the agents. As depicted in Fig. 5, the agents must navigate around obstacles within the environment, with both the number and the placement of obstacles differing across the various configurations. Subfigures (a), (b), and (c) demonstrate different configurations of agents, obstacles, and target positions within the task. The colored circles represent different components within the scenario: blue, red, and yellow circles denote Agent 1, Agent 2, and Agent 3, respectively; the green circle represents the box being pushed; the black circles indicate obstacles; and the gray circle marks the target position. The positions and sizes of the components in the box-pushing task scenario are detailed in Table 3.

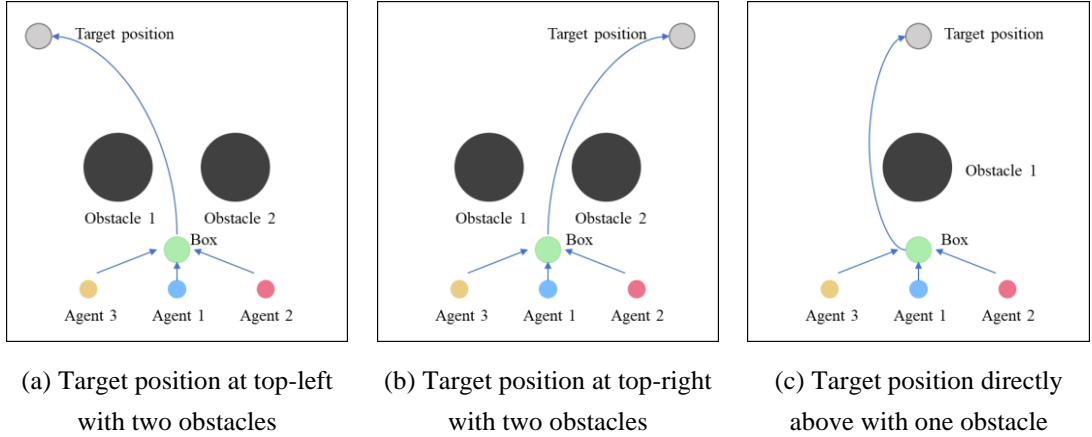

(a) Target position at top-left with two obstacles
(b) Target position at top-right with two obstacles
(c) Target position directly above with one obstacle

Fig. 5 The box-pushing task scenario

Table 3. Detailed settings of the box-pushing task scenario

| Component | Setting |
| --- | --- |
| Agent 1 | Position: (0, -0.75); Size: 0.05 |
| Agent 2 | Position: (0.5, -0.75); Size: 0.05 |
| Agent 3 | Position: (-0.5, -0.75); Size: 0.05 |
| Box | Position: (0, -0.5); Size: 0.075 |
| Obstacle 1 | Position: Fig. 5 (a) and (b): (-0.3, 0), Fig. 5 (c): (0, 0); Size: 0.2 |
| Obstacle 2 | Position: (0.3, 0); Size: 0.2 |
| Target Position | Position: Fig. 5 (a): (-0.9, 0.9), Fig. 5 (b): (0.9, 0.9), Fig. 5 (c): (0, 0.9); Size: 0.075 |

Note: The positions are given in absolute coordinates with the center of the task area at (0,0), and the sizes correspond to the radius of the circles.

## 4.2 Reward Function

In this paper, the reward function is designed with reference to the methods outlined in [14, 59].

The specific configuration of the reward function is described as follows.

**1) Distance reward.** A reward is given at each step based on distance change: positive when the distance of box to the target position decreases, negative when it increases, with the reward magnitude proportional to the change in distance. $D_{t-1}$ and $D_t$ represent the Euclidean distances between the box and target position at the previous and current timesteps, respectively.

$$R_{dis} = (D_{t-1} - D_t) \times 50 \tag{10}$$

**2) Push reward.** A positive reward is given when the agent pushes the box. Specifically, a reward is provided when the agent performs the action of pushing the box.

$$R_{push} = \begin{cases} 50 & \text{if push occurs} \\ 0 & \text{if no push occurs} \end{cases} \tag{11}$$

**3) Goal reward.** A significant reward is granted when the box reaches the target position.

$$R_{goal} = \begin{cases} 1000 & \text{if target position is reached} \\ 0 & \text{if target position is not reached} \end{cases} \tag{12}$$

**4) Collision reward.** A negative reward is given if a collision occurs either between the agents or between the box and an obstacle.

$$R_{col} = \begin{cases} -50 & \text{if collision occurs} \\ 0 & \text{if no collision occurs} \end{cases} \tag{13}$$

**5) Boundary reward.** A negative reward is imposed if the agent exceeds the boundary.

$$R_{bound} = \begin{cases} -50 & \text{if boundary is exceeded} \\ 0 & \text{if boundary is not exceeded} \end{cases} \tag{14}$$

The total reward is the sum of all these individual rewards, as expressed in Eq. (15).

$$R_{total} = R_{dis} + R_{push} + R_{goal} + R_{col} + R_{bound} \tag{15}$$

It should be noted that, as our focus is on investigating the emergence of hierarchical structures among agents pursuing a joint objective, our reward function is specifically designed to incentivize individual agent behaviors rather than explicitly promoting collaborative actions or pre-defined role allocations.

### 4.3 Observation and Action Spaces

Central to the CTDE framework in Fig. 1 is the definition of each agent's state and their admissible action. Consequently, we shall proceed by formally defining the observation and action spaces for the agents in the box-pushing exercise.

**4.3.1 Observation Space**

At each time step, the agents obtain local observations, which can be classified into four categories of variables: agent-related variables, environment-related variables, task-related variables, and team-related variables (as defined in Section 3.1). These components are then combined to form the $i^{th}$, $i = 1, \dots, n$ agent's observation space $O_i$, as represented in Eq. (16).

$$O_i = \{o_{i,1}, o_{i,2}, \dots, o_{i,d_i}\}, \quad i = 1, \dots, n, \tag{16}$$

where $d_i$ denotes the dimensionality of the $i^{th}$ agent's observation space.

**1) Agent-related variables.**
- **Position.** The current position of the agent in the environment, including its coordinates in the $x$ and $y$ directions.
- **Velocity.** The current velocity of the agent, including its motion rates in the $x$ and $y$ directions within the environment.

**2) Environment-related variables.**

- **Relative position of obstacles.** The relative position of obstacles with respect to the agent, including the $x$ and $y$ directional distances, which helps the agent avoid obstacles.

**3) Task-related variables.**

- **Relative position of the agent to the target position.** The relative position of the agent to the target, including the $x$ and $y$ directional distances, which enables the agent to assess its distance from the target and plan the appropriate movement.
- **Relative position of the box to the target position.** The relative position of the box to the target, including the $x$ and $y$ directional distances, which provides the agent with the proximity of the box to the target.

**4) Team-related variables.**

- **Position of other agents.** The positions of other agents within the team, including their coordinates in the $x$ and $y$ directions, which are essential for coordination and collaboration.
- **Velocity of other agents.** The velocities of other agents, including their speeds in the $x$ and $y$ directions, which allow the agent to understand the dynamics of its teammates and adjust its actions accordingly.

In Fig. 5, the number of obstacles differs across the three scenarios of the box-pushing task, resulting in variations in the observation space. Specifically, Fig. 5 (a) and (b) each contain two obstacles, whereas Fig. 5 (c) includes only one obstacle. The corresponding observation spaces for these scenarios are represented by 20-dimensional and 18-dimensional vectors, respectively.

**4.3.2 Action Space**

In the MPE, the original action space for the particles is continuous. To simplify the problem and improve computational efficiency, we discretize the action space by defining the agents' actions as movements in specific directions. The action space for the $n$ agents include moving left, right, down, up, or remaining stationary, is given as

$$A_i = \{a_{i,1}, a_{i,2}, a_{i,3}, a_{i,4}, a_{i,5}\}, \ i = 1, \dots, n \tag{17}$$

where $a_1$ represents moving left, $a_2$ represents moving right, $a_3$ represents moving down, $a_4$ represents moving up, and $a_5$ represents remaining stationary.

To further clarify, consider a scenario where the agent selects one of these directions based on the current task. In the MPE, we represent the agent's action as a binary selection, where a value of 1 indicates the chosen action, and a value of 0 indicates all other unchosen actions. For example, if the agent selects moving up, the action space would be represented as $\{0,0,0,1,0\}$.

**4.4 Hyperparameter Settings**

The hyperparameter settings for the MADDPG algorithm utilized in this paper are summarized in Table 4. These choices were made to balance training stability, convergence speed, and computational resources based on the algorithm's requirements in the MPE.

Table 4. Hyperparameter settings for the MADDPG algorithm

| Hyperparameter | Value | Description |
| --- | --- | --- |
| Maximum episode length | 50 | Maximum number of time steps allowed for each episode |
| Maximum episodes | 20,000 | Maximum number of episodes to run |
| Learning start step | 50,000 | Number of steps before learning begins |
| Learning frequency | 100 | Number of time steps between each learning update |
| Max gradient norm | 0.5 | Maximum gradient norm for clipping |
| Exchange depth ($\tau$) | 0.01 | Depth of parameter exchange in the neural network |

| Learning rate (actor) | 0.01 | Learning rate for the actor optimizer |
| Learning rate (critic) | 0.01 | Learning rate for the critic optimizer |
| Discount factor ($\gamma$) | 0.97 | Discount factor for future rewards |
| Batch size | 1,256 | Number of episodes used for each optimization step |
| Memory size | 100,000 | Number of stored data points in memory |
| MLP units (layer 1) | 128 | Number of units in the first hidden layer of MLP |
| MLP units (layer 2) | 64 | Number of units in the second hidden layer of MLP |

## 5 Results and Discussion

In this section, we train our MASOS using the MADDPG algorithm for the box-pushing exercise. Following successful training, we conduct step-by-step analysis of the model's task execution across various testing conditions to investigate its emergent behaviors. Specifically, we investigate whether hierarchical structures emerge during task execution and explore how task environment configurations and network initialization conditions influence the emergence of such hierarchies.

### 5.1 Emergence of Dependency Hierarchy

Initially, MASOS are trained to perform the box-pushing exercise in the scenario illustrated in Fig. 5 (a). The variation in the total reward over 20,000 episodes during the training process is shown in Fig. 6. The $x$-axis represents the number of episodes, while the $y$-axis corresponds to the cumulative reward obtained by the agent team over each training episode. A training episode is one complete box-pushing exercise from initial position to the target position). In Fig. 6, the original results are depicted by the light red curve, and the smoothed results, obtained by applying a moving average with a window size of 100, are represented by the blue curve. The total episode reward consistently increases with the number of training episodes and eventually stabilizes within a fixed range. This indicates that the agent team, through training with the MADDPG algorithm, has successfully learned an action policy aimed at maximizing cumulative rewards.

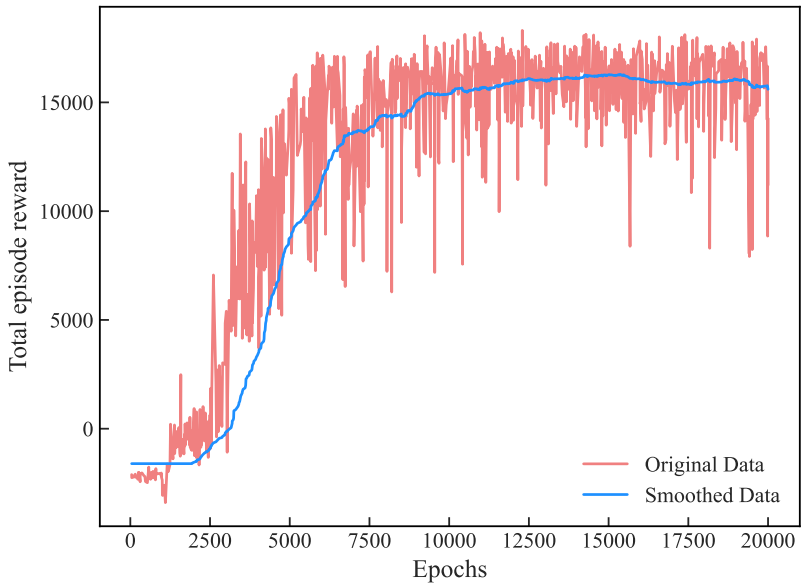

Fig. 6 Total episode reward over epochs

Subsequently, MASOS employ the trained model to execute the box-pushing task, and a comprehensive retrospective analysis of the entire task process is conducted, focusing on the sensitivity of each agent's actions to the states of other agents and the dynamic changes in each agent's dependency values. The sensitivity of each agent's action with respect to the states of other agents is computed using Eq. (7). The resulting sensitivity curve over 50 time steps (from step 0 to step 49) during task execution is illustrated in Fig. 7. The six curves respectively depict the sensitivity of Agent 1's actions to the states of Agents 2 and 3, Agent 2's actions to the states of Agents 1 and 3, and Agent 3's actions to the states of Agents 1 and 2 during task execution. Following this, the dependency value of each agent is calculated using Eq. (8), and the corresponding dependency value curves during task execution are presented in Fig. 8. At each time step, these dependency curves sum to zero, consistent with the calculation in Eq. (9). This indicates that within the overall system, the interplay between each agent's dependency and reverse dependency keeps the total dependency sum of all agents stable. However, the dependency values of each agent dynamically change at different time steps, with some agents exhibiting the highest dependency values at certain points. It is observed that Agent 1 exhibits the highest dependency value from steps 0 to 11, while Agent 2 demonstrates the highest dependency value between steps 12 and 37. From steps 38 to 49, Agent 1 again shows the highest dependency value. The visualization of these steps during task execution is shown in Fig. 9. Specifically, steps 0 to 11 correspond to initially pushing the box upward until it encounters an obstacle; steps 12 to 37 correspond to maneuvering the box through a narrow path between two obstacles; and steps 38 to 49 correspond to pushing the box directly toward the target position.

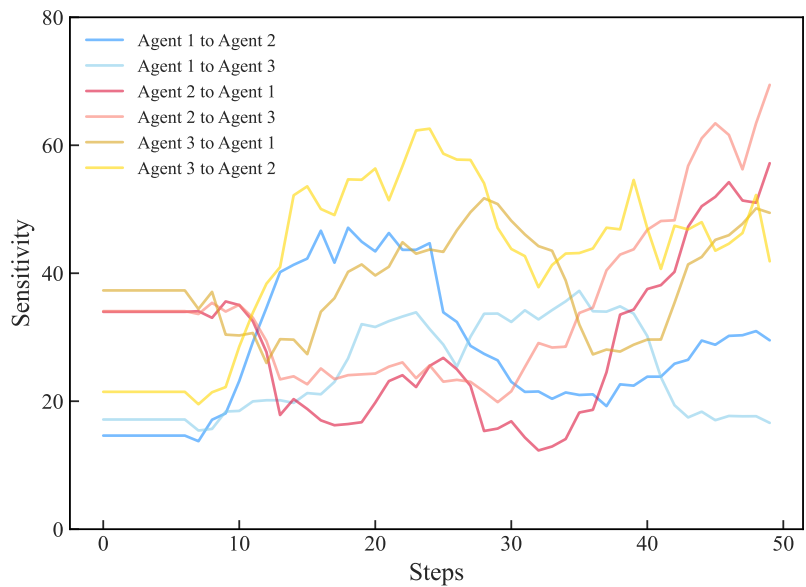

Fig. 7 Sensitivity curve (target position at top-left with two obstacles)

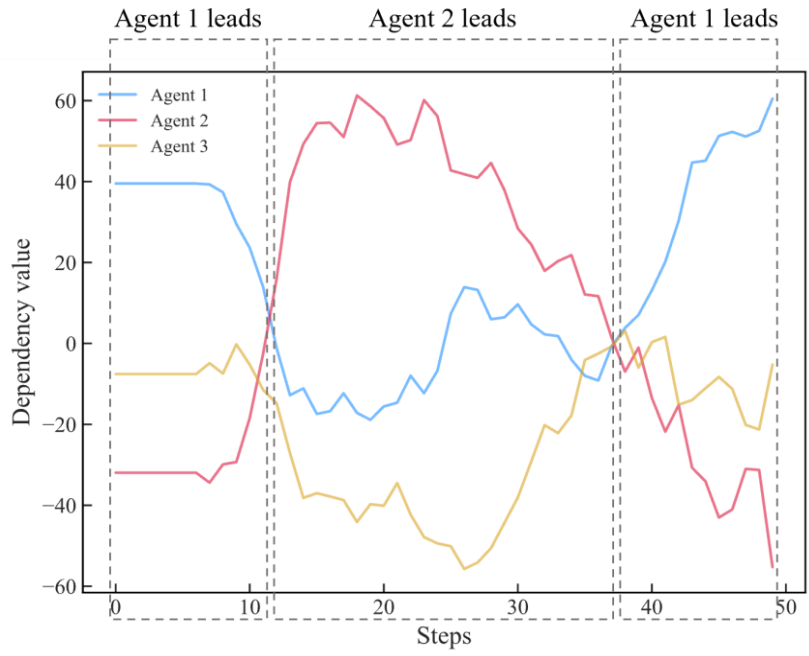

Fig. 8 Dependency value curve (target position at top-left with two obstacles)

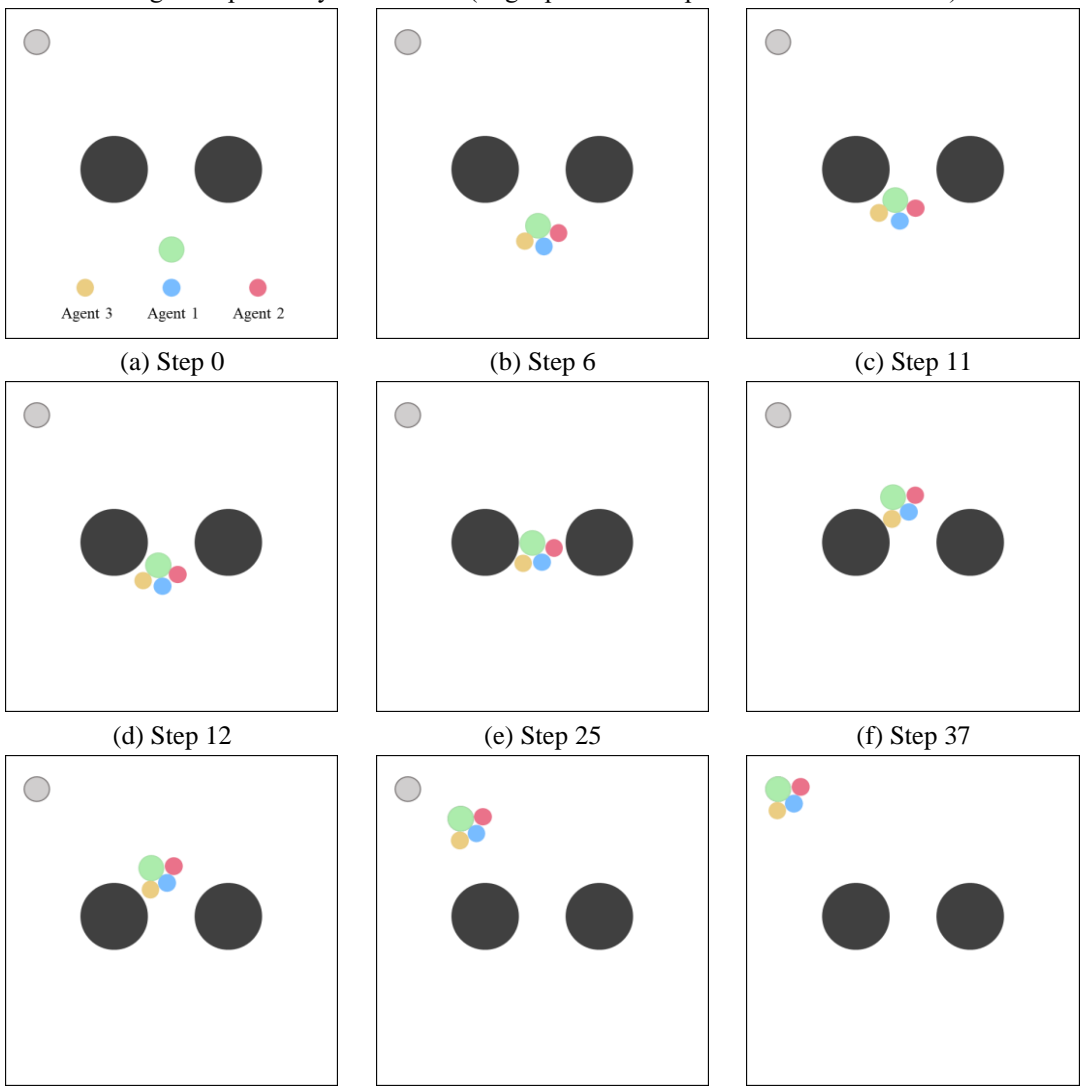

(a) Step 0  (b) Step 6  (c) Step 11

(d) Step 12  (e) Step 25  (f) Step 37

(g) Step 38                    (h) Step 44                    (i) Step 49

Fig. 9 Visualization of key timesteps in task execution (target position at top-left with two obstacles)

A plausible explanation for the observed behavioral pattern is that during the initial task phase, Agent 1 exhibited the highest dependency value within the team, guiding the collective effort to push the box upward. This phenomenon can be attributed to Agent 1's positional advantage, as its initial proximity to the box resulted in its actions being most heavily relied upon by other agents. As the task progressed, requiring the box to be maneuvered through a narrow path between two obstacles followed by a left turn to circumvent an obstruction, Agent 2 emerged as the agent with the highest dependency value. This shift likely reflects its more critical role in facilitating the rotational manipulation of the box. Finally, as the task approached completion and the box required direct propulsion toward the target position, Agent 1 regained its status as the most dependent-upon agent, orchestrating the team's final push to the goal.

Based on these findings, we observe the emergence of a dependency hierarchy in MASOS during task execution, which dynamically adapts to the evolving demands of the task. Specifically, Agent 1 dominates during linear pushing phases due to its positional advantage, while Agent 2 assumes greater dependency centrality during rotational maneuvers. This adaptive hierarchy enhances the system's efficiency, robustness, and ability to handle complex tasks, illustrating the potential of MASOS to achieve collective intelligence through self-organization.

## 5.2 Effect of Task Environments

To explore the effect of the task environment on the emergence of dependency hierarchy, we relocated the target position from the top-left to the top-right, as shown in Fig. 5 (b). The dependency value curves of the agents during task execution are presented in Fig. 10, with key steps illustrated in Fig. 11. The results demonstrate that Agent 1 exhibits the highest dependency value during the initial task stage (steps 0-12), when pushing the box upward until encountering an obstacle. Subsequently, Agent 3 shows the highest dependency value during the intermediate phase (steps 13-28), which involves navigating the box through a narrow passage between two obstacles followed by a right turn to bypass them. Finally, Agent 1 regains the highest dependency value during the terminal phase (steps 29-49), when directly pushing the box toward the target position. These findings confirm the dynamic emergence of dependency hierarchy in MASOS during task execution.

Furthermore, comparative analysis between Fig. 8 and Fig. 10 reveals distinct dependency patterns: Agent 2 dominates during leftward rotations (Fig. 8), while Agent 3 prevails during rightward rotations (Fig. 10). This suggests that Agent 2 plays a pivotal role in left-turn maneuvers, whereas Agent 3 is more critical for right-turn maneuvers. These results highlight the significant influence of target position on the emergence of dependency hierarchies in MASOS. More importantly, MASOS exhibit environment-dependent emergence of distinct hierarchical structures, where agents dynamically adapt their roles to optimize collective performance.

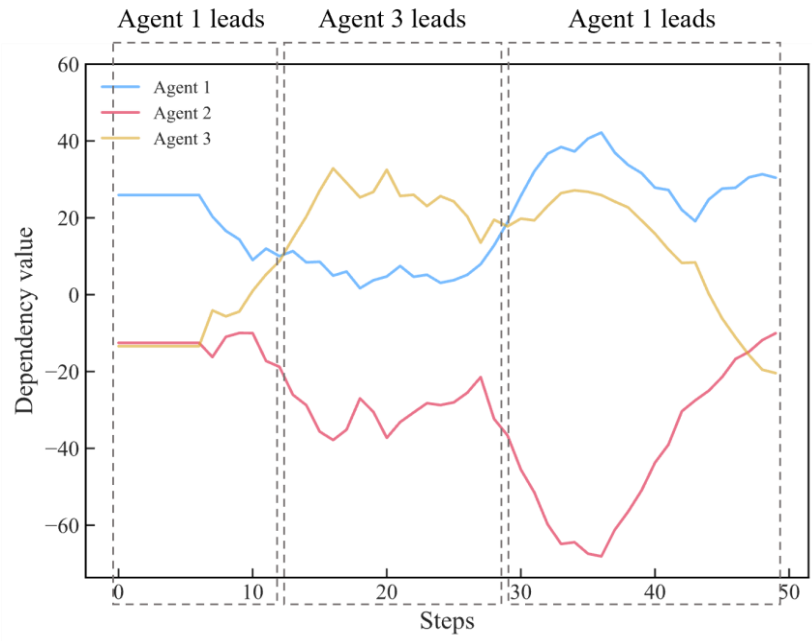

Fig. 10 Dependency value curve (target position at top-right with two obstacles)

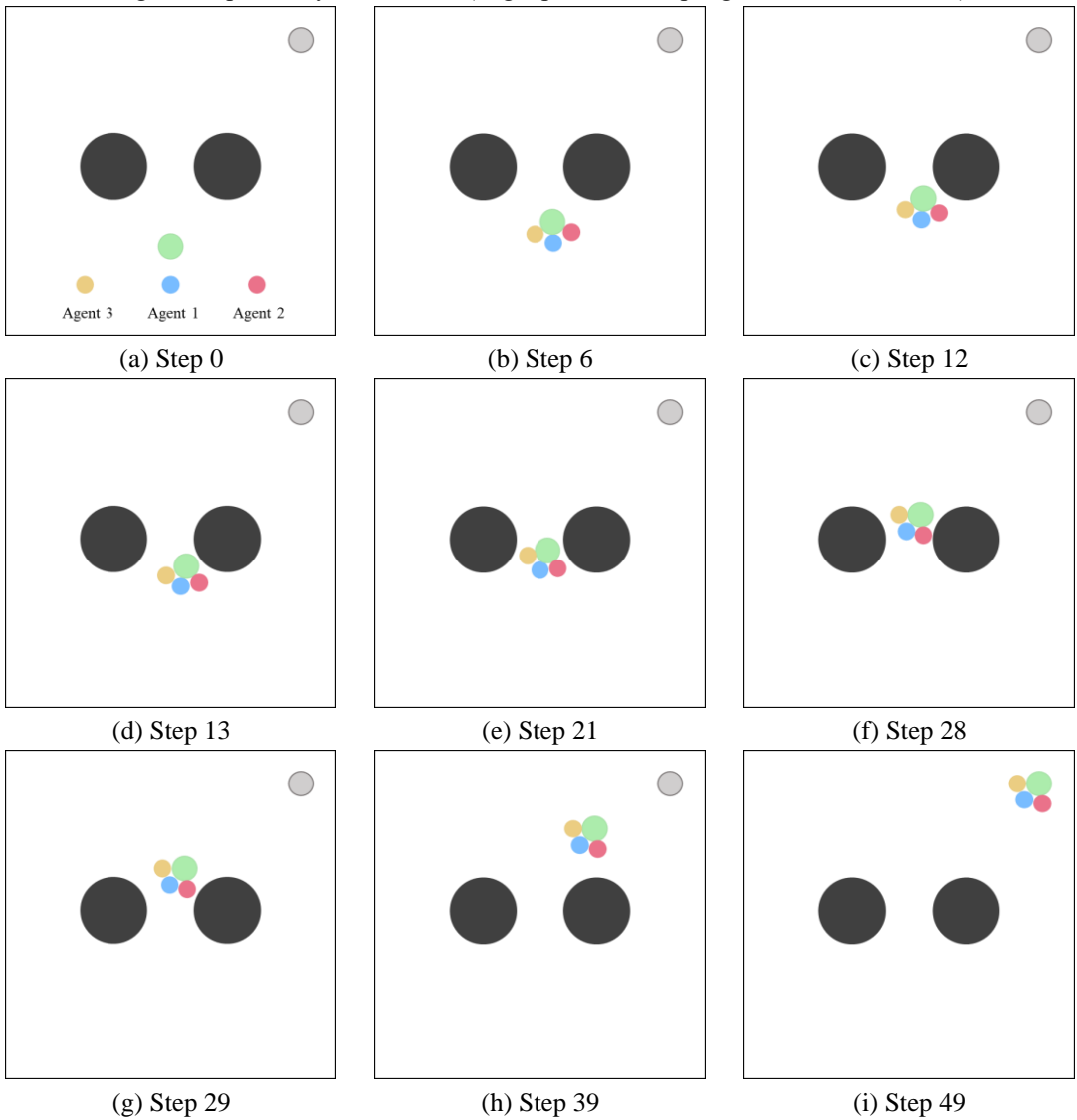

(a) Step 0    (b) Step 6    (c) Step 12

(d) Step 13   (e) Step 21   (f) Step 28

(g) Step 29   (h) Step 39   (i) Step 49

Fig. 11 Visualization of key timesteps in task execution (target position at top-right with two obstacles)

To further investigate this environmental influence, we analyze a configuration with the target positioned directly above a single obstacle, as illustrated in Fig. 5 (c). In this setup, the trained MASOS can navigate the box to the target by circumventing the obstacle either leftward or rightward. The corresponding dependency value curves for left and right bypass trajectories, which are obtained by training the model from scratch, are presented in Fig. 12 (a) and (b), respectively. Key execution steps for these two cases are illustrated in Fig. 13 and Fig. 14. The results reveal a three-phase dependency pattern: (1) During initial upward pushing, Agent 1 consistently demonstrates the highest dependency value; (2) In the obstacle circumvention phase, Agent 3 exhibits peak dependency during left bypass trajectories (Fig. 12 (a)), while Agent 2 shows maximum dependency during right bypass maneuvers (Fig. 12 (b)); (3) Finally, Agent 1 regains dominance during the terminal target approach. Notably, the left-bypass hierarchy (Fig. 12 (a)) mirrors the pattern observed in Fig. 10 (target position at top-right), whereas the right-bypass hierarchy (Fig. 12 (b)) replicates the Fig. 8 configuration (target position at top-left).

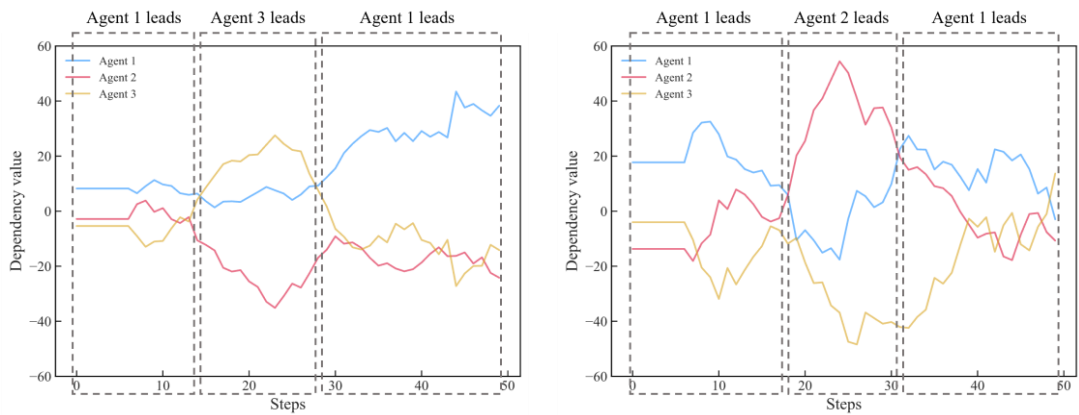

(a) Bypassing obstacle from the left  (b) Bypassing obstacle from the right

Fig. 12 Dependency value curve (target position directly above with one obstacle)

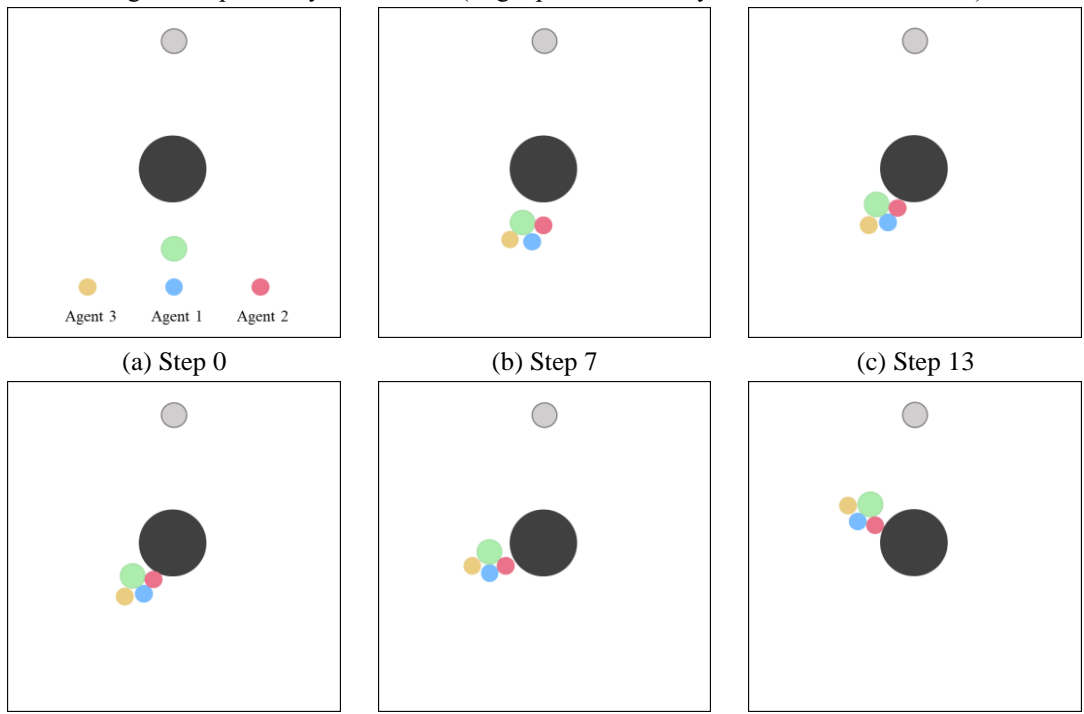

(a) Step 0  (b) Step 7  (c) Step 13

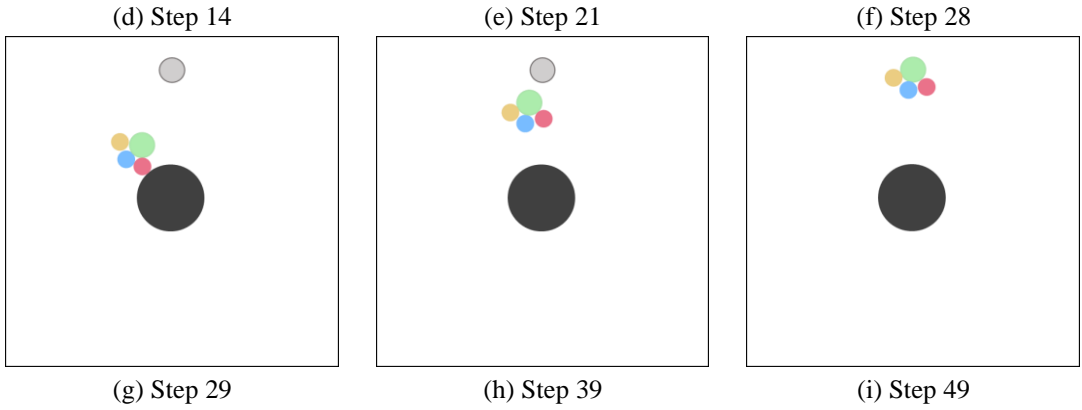

(d) Step 14    (e) Step 21    (f) Step 28

(g) Step 29    (h) Step 39    (i) Step 49

Fig. 13 Visualization of key timesteps in task execution: bypassing obstacle from the left (target position directly above with one obstacle)

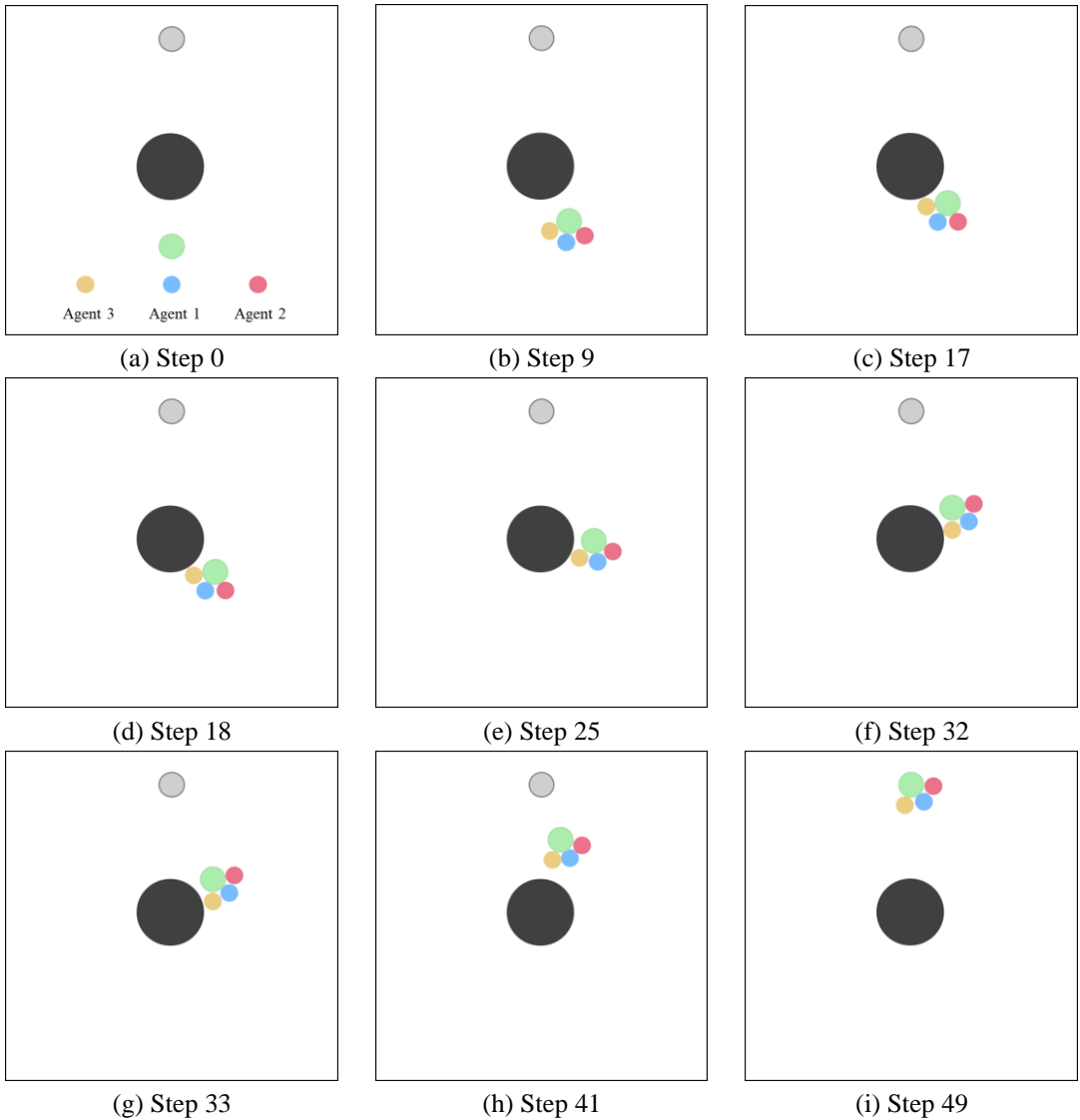

(a) Step 0    (b) Step 9    (c) Step 17

(d) Step 18    (e) Step 25    (f) Step 32

(g) Step 33    (h) Step 41    (i) Step 49

Fig. 14 Visualization of key timesteps in task execution: bypassing obstacle from the right (target position directly above with one obstacle)

Our results demonstrate the emergence of clear role specialization within MASOS. Agent 1 consistently achieves peak dependency values during linear pushing phases, thereby establishing leadership in straightforward navigation tasks. In contrast, Agents 2 and 3 emerge as critical controllers

Page 25

during turning maneuvers, with their relative importance showing strong direction-dependence. Specifically, circumventing the obstacle from the left significantly increases Agent 3's dependency values, while bypassing the obstacle from the right preferentially enhances Agent 2's dependency metrics. This directional specialization can be attributed to their inherent spatial configurations. Agent 3's left-biased positional advantage enables it to maximize informational influence during leftward maneuvers. Similarly, Agent 2's right-sided location allows it to optimize steering contribution during rightward maneuvers.

These findings not only demonstrate the adaptability of MASOS to varying task environments, but also reveal its environment-dependent emergence of hierarchical structures. Agents with positional advantages attain the highest dependency values, thereby exerting primary influence on team decision-making processes. In contrast, other agents dynamically adjust their behaviors in response to these spatially determined leaders. More fundamentally, the emergent hierarchies result from continuous interactions between agent positioning, task demands, and environmental configurations. This adaptive mechanism enables MASOS to efficiently accomplish complex objectives through self-organized cooperation, with role specialization emerging naturally from contextual requirements rather than being pre-programmed.

In addition, it is observed that the MASOS, trained with random initialization, exhibits varying dependency hierarchies during task execution (e.g., Fig. 12). To further investigate the extent to which this randomness influences the emergence of hierarchies, it is essential to consider the role of the random initialization of the initial policy function networks.

**5.3 Effect of Network Initialization Conditions**

Network initialization conditions significantly influence agent behavior by determining initial policy parameters. To systematically investigate their impact on the emergence of dependency hierarchies, we conduct experiments using six distinct random seeds (seed = 5, 10, 15, 20, 25, 30) across all three task configurations shown in Fig. 5. The resulting dependency value curves are presented in Figs 14 to 16, which correspond to the scenarios with: target position at top-left with two obstacles (Fig. 15), target position at top-right with two obstacles (Fig. 16), and target position directly above with one obstacle (Fig. 17), respectively.

As illustrated in Fig. 15, varying random seeds yield distinct shapes of dependency value curves. In certain cases (Fig. 15 (b), (d), (f)), Agent 1 consistently maintains the highest dependency value throughout task execution, indicating its sustained dominance in team decision-making. Conversely, other cases (Fig. 15 (a), (c), (e)) exhibit phase-dependent leadership transitions, where different agents achieve dominance during specific task phases. Notably, these two characteristic patterns of dependency value curves are also observed in Fig. 16 and Fig. 17. The results demonstrate that across all three task scenarios, different network initialization conditions yield distinct yet characteristic dependency value curves, revealing two primary hierarchy emergence patterns: (1) persistent dominance, where a single agent maintains the highest dependency value throughout the entire task execution, continuously guiding team decisions; and (2) alternating dominance, where leadership dynamically shifts between agents during different task stages (e.g., Agent 1 dominating linear pushing phases while Agents 2 or 3 lead turning maneuvers). Finally, it is worth nothing that when Agent 1 exhibits a persistent dominance pattern, we consistently observe an inverse dependency relation between Agents 2 and 3 during the rotational maneuver.

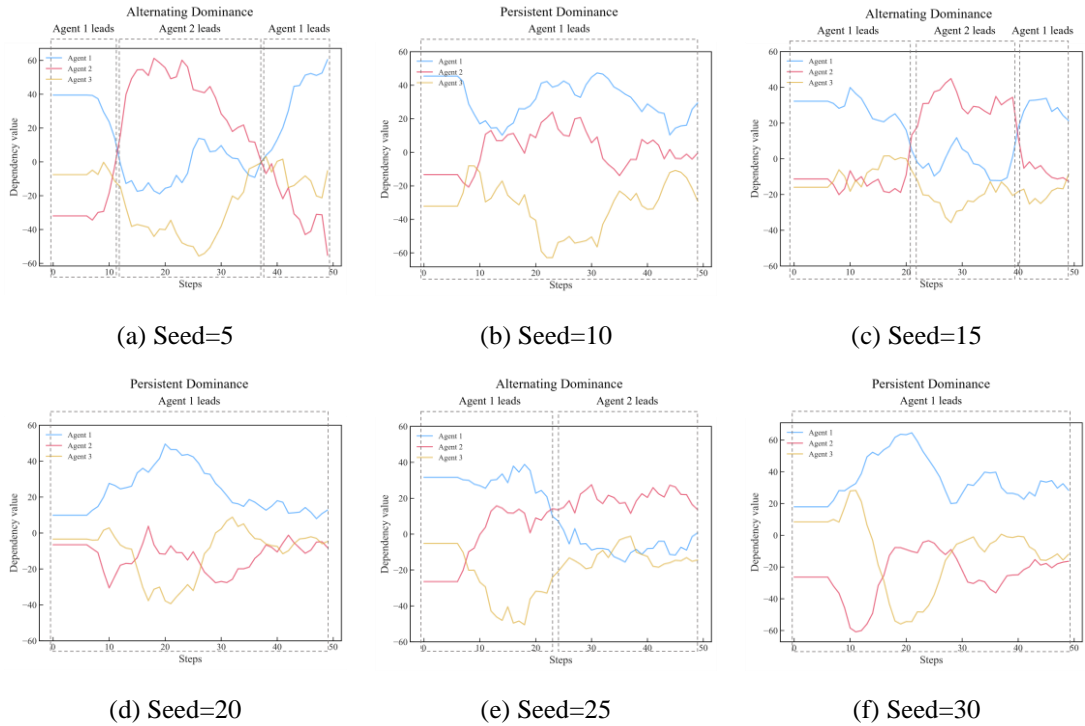

Fig. 15 Dependency value curves under different network initialization conditions (target position at top-left with two obstacles)

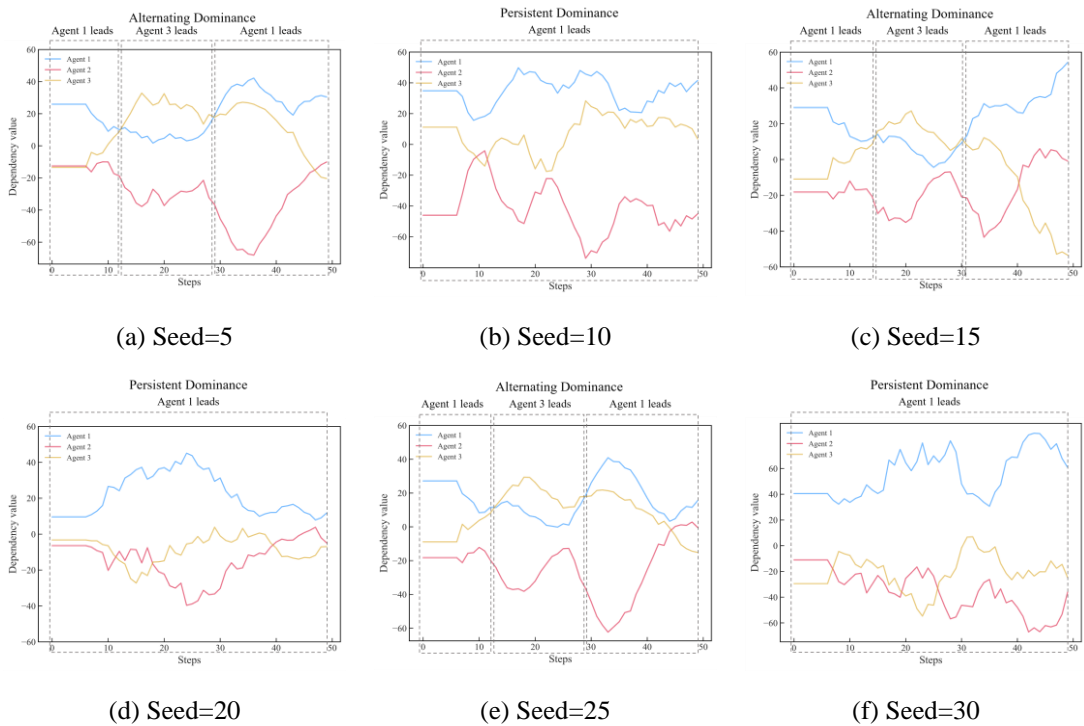

Fig. 16 Dependency value curves under different network initialization conditions (target position at top-right with two obstacles)

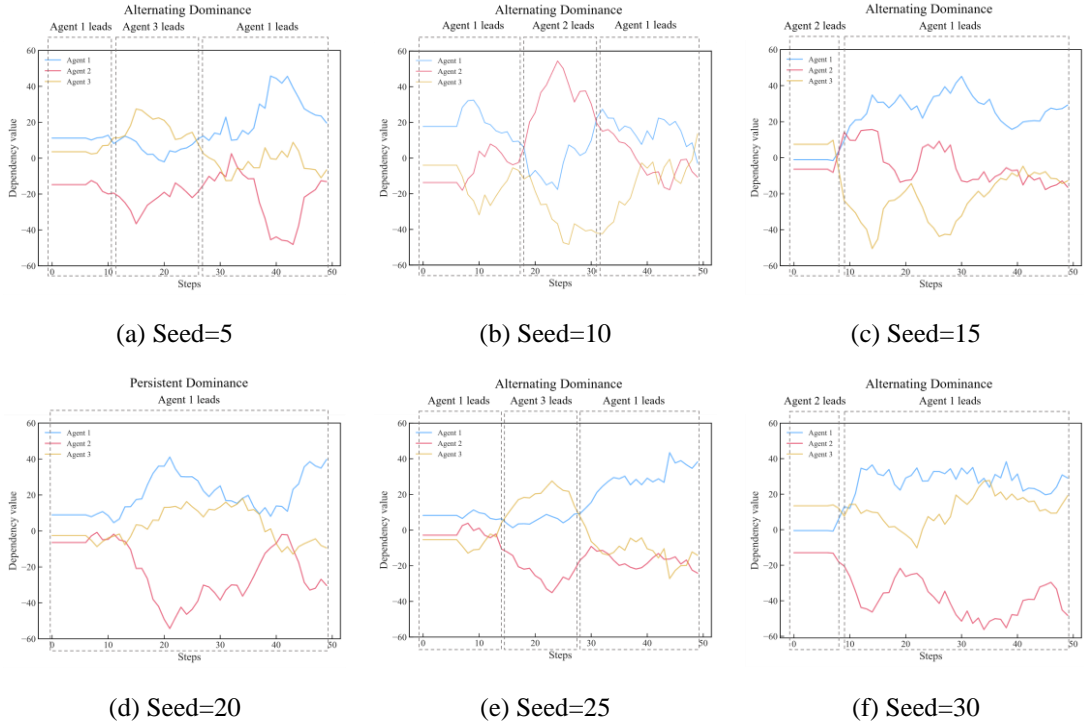

Fig. 17 Dependency value curves under different network initialization conditions (target position directly above with one obstacle)

These results collectively demonstrate two key aspects of MASOS: (1) the consistent emergence of dependency hierarchies across varying network initialization conditions during task execution, and (2) the significant influence of network initialization conditions on hierarchical formation patterns. Under different experimental configurations, the system exhibits two distinct types of emergent hierarchies - persistent dominance hierarchies (where a single agent maintains continuous leadership) and alternating dominance hierarchies (characterized by phase-dependent role transitions).

## 5.4 Discuss about Talent, Environment, and Effort

In the previous sections, we analyzed how MASOS give rise to dependency hierarchies during task execution, and the effect of task environments (such as agents' positional advantages) and network initialization conditions on the emergence of these hierarchies. Variations in the task environment (e.g., adjustments to the target position) and differences in network initialization conditions influence an agents' relative dependency curve, resulting in the formation of distinct hierarchical structures. Specifically, changes in the task environment mainly alter agents' positional advantages, while differing network initialization conditions affect the agents' action outputs. Based on these observations, we now delve deeper into the concepts of "Talent," "Environment," and "Effort."

"Environment" represents the external conditions of MASOS, including task configurations such as the initial position of the box, the target position, and the location of obstacles. "Talent" refers to inherent advantages determined before task execution—such as favorable positioning or optimal network initialization—that allow an agent to perform more effectively under certain circumstances. In contrast, "Effort" represents the agents' learning process, which corresponds to modifications in their behavior and policies over time through interactions with the "Environment." "Effort" reflects the improvements or adaptations an agent makes during task execution, demonstrated through its ability to adjust to environmental changes and optimize its actions.

The dependency value of each agent can be viewed as a dynamic interaction between "Talent" and "Effort" within the "Environment." Agents with favorable initial conditions—those possessing greater "Talent"—may initially attain higher dependency values and consequently assume leadership roles. However, the "Effort" expended by agents during task execution, reflected by updates in their policies through learning, enables them to shift roles, increase their influence, and contribute substantially to the system's overall performance. For example, in Fig. 8, the dependency value of Agent 2 is lowest during steps 0 to 11, which corresponds to the "pushing the box upward" phase. However, through the interaction with the "Environment" and the updates in its policy, Agent 2 shows the highest dependency value during steps 12 to 37, which corresponds to the "maneuvering the box through a narrow path between two obstacles" phase. This demonstrates that Agent 2 transitions from a subordinate role to a leadership role through "Effort." Similarly, in Fig. 10, Agent 3 shifts from a subordinate role during steps 0 to 12 to a leadership role during steps 13 to 28. This dynamic relationship between "Talent" and "Effort" within the "Environment" allows MASOS to self-optimize and adjust, adapting to real-time task demands. A crucial aspect of this relationship is that "Effort" does not merely compensate for a lack of "Talent"; instead, it complements and enhances an agent's initial position. An agent with relatively low "Talent" may still exert significant influence if it continuously adapts and refines its policies through learning. Conversely, an agent with strong "Talent" but little adaptation through "Effort" may fail to fully exploit its advantages.

During certain phases of the task, agents with stronger "Talent" may dominate task execution; however, as the task progresses, agents with greater "Effort" may shift the balance of positions, altering the hierarchical roles. This continuously evolving hierarchy showcases the system's self-organizing capacity, enabling MASOS to optimize in response to the changing task requirements. While initial advantages guide early decision-making and task execution, the ongoing learning and interaction among agents allow the system to progressively improve. This adaptability is critical in complex, dynamic, and uncertain environments, particularly when dealing with increasing task complexity and the need for real-time learning.

"Talent" determines the initial role, but through continuous "Effort" within the "Environment," agents can alter their roles and positions within the system. The interaction between "Talent" and "Effort" within the "Environment" not only influences the emergence of hierarchies but also affects the long-term stability and resilience of MASOS. Systems that rely too heavily on initial "Talent" may experience stagnation or rigidity, especially in the dynamic "Environment." Conversely, systems that encourage sustained "Effort" from all agents maintain flexibility, enabling agents to adjust their roles and contribute to the realization of the system's joint objectives as the task evolves.

In summary, the balance between "Talent" and "Effort" within the "Environment" of MASOS leads to a dynamic, flexible, and adaptive hierarchy that evolves throughout task execution. This interplay allows MASOS to leverage the agents' initial advantages, promoting the emergence of collective intelligence while enabling agents to adapt to the complexities and uncertainties of real-world scenarios. Understanding and optimizing this relationship is essential for the design and regulation of MASOS, particularly in larger-scale and more complex applications. It is important to note that these findings, which consider a team of agents working towards a joint objective, may not generalize to systems of agents with disparate or ulterior objectives.

# 6 Conclusion and Future Work

In this paper, we addressed the following question:

*Do hierarchies emerge in teams of agents trained for a joint objective?*

To address this, MARL is employed to train MASOS for a collaborative box-pushing task. The inter-agent dependencies are quantified by calculating the gradients of each agent's actions relative to the states of other agents. The emergence of hierarchies is then analyzed through the aggregation of these dependencies. The findings of our study include the following:

(1) Hierarchical structures emerge within MASOS during the execution of a joint objective task. These hierarchies are dynamic, adjusting in response to the changing demands of the task. As the task progresses, the hierarchy evolves based on the agents' positional advantages and the specific phases of the task, such as linear pushing and rotational maneuvers. This adaptability enhances the collective intelligence of the system, as agents shift roles and optimize their contributions to meet real-time task requirements.

(2) The emergence of hierarchies in MASOS is significantly influenced by the task environment and network initialization conditions. Under varying task environments and network initialization settings, two distinct hierarchical structures are observed: persistent dominance, where a single agent maintains leadership throughout the task, and alternating dominance, where leadership roles shift according to the task phases.

(3) The emergence of hierarchies in MASOS arises from the dynamic interplay between agents' "Talent" and "Effort" within the "Environment." "Talent" refers to an agent's inherent advantages, while "Effort" represents its learning process. While "Talent" establishes the agent's initial role, continuous "Effort" within the "Environment" enables agents to alter their roles and positions within the system. The interaction between these factors determines the agent's influence on team decision-making, allowing MASOS to develop a dynamically evolving hierarchy during task execution.

The box-pushing problem serves as a representative example of an MASOS, reflecting a range of complex practical challenges, including multi-robot collaborative assembly in manufacturing, multi-robot rescue planning in disaster relief, and multi-robot task and path planning in industrial logistics. The findings from this paper provide valuable insights into understanding self-organizing behaviors in MASOS and offer guidance for practical applications.

While the box-pushing problem offers valuable insights, it remains a simplified model compared to more complex practical problems. We also acknowledge that the MASOS examined in this paper are relatively small in scale, involving only a few agents. Future research will focus on increasing the number of agents and expanding the range of admissible actions in the task. Additionally, we aim to explore how these findings can be applied to a broader range of practical scenarios and investigate methods for regulating and optimizing emergent behaviors in MASOS.

## Acknowledgment

Gang Chen acknowledges support from China Scholarship Council (202506030002). Zhenjun Ming acknowledges support from the National Natural Science Foundation of China (62373047), and the Beijing Municipal Science and Technology Foundation (3222020). Guoxin Wang acknowledges support from the National Natural Science Foundation of China (51975056).